\documentclass[aps,pra,reprint,twocolumn,groupedaddress]{revtex4-2}
\usepackage{amsmath}
\usepackage{esint}
\usepackage{graphicx}
\usepackage{xcolor}
\usepackage{siunitx}

\newcommand{\REMPIM}{\ensuremath{M}}
\newcommand{\REMPIN}{\ensuremath{N}}

\begin{document}


\title{Measurement of ionization delays in atomic REMPI using photoelectron vortices}


\author{D. Köhnke, T. Bayer and M. Wollenhaupt}
\affiliation{Carl von Ossietzky Universität Oldenburg, Institut für Physik, Carl-von-Ossietzky-Straße 9-11, D-26129 Oldenburg, Germany}


\date{\today}

\begin{abstract}
We study time-delays in atomic multiphoton ionization (MPI) using the phase-sensitive detection of photoelectron vortices. Photoelectron vortices are created by MPI with counter-rotating circularly polarized (CRCP) ultrashort double pulse sequences. The vortices are characterized by a spiral-shaped photoelectron momentum distribution (PMD) resulting from the Ramsey-type interference of the partial photoelectron wave packets created by each pulse. 
The slope of the spiral arms encodes the time-delay between the partial wave packets with interferometric precision.
We study the ionization time-delay in the (1+2) resonance-enhanced multiphoton ionization (REMPI) of potassium atoms using femtosecond CRCP white-light supercontinuum pulses. Pairwise interference of the partial wave packets with a reference wave packet gives rise to vortices of different rotational symmetry, which are superimposed to form the total PMD. The PMD is reconstructed tomographically and decomposed into the vortices using 3D Fourier analysis. From the energy-dependent spectral phases of the vortices, we retrieve a resonant ionization delay in the order of \SI{1}{fs}, much shorter than the pulse duration. Our results demonstrate the power of photoelectron vortices for the accurate determination of photoionization time-delays and open up new perspectives for photoelectron chronoscopy including the measurement of ionization delays on the attosecond timescale.
\end{abstract}


\maketitle


\section{Introduction\label{sec:Introduction}}
The photoelectric effect is one of the fundamental processes of light-matter interaction. Its interpretation by A. Einstein \cite{Einstein:1905:AP:165} made a major contribution to the development of quantum physics. Today, photoelectron spectroscopy has become a powerful technique for studying light-induced quantum dynamics in atoms, molecules and condensed phase systems. The fundamental question of how long photoemission takes has been the subject of a long-standing debate, recently fueled by advances in attosecond metrology \cite{Hentschel:2001:Nature:509,Gallmann:2012:ARPC:447,Hockett:2018b} and new capabilities to directly observe ultrafast electron dynamics in real time (see e.g. \cite{Corkum:2007:NP:381,Kling:2008:ARPC:463,Krausz:2009:RMP:163,Gallmann:2012:ARPC:447,Calegari:2016:JPB:062001,Borrego-Varillas:2022:RPP:066401} and references therein). Different types of time-delays have been identified in photoionization. In general, a distinction is made between time-delays acquired by the electron during its transition from the initial state to the continuum and those subsequently accumulated during the propagation of the outgoing photoelectron in the continuum. A prominent example related to the propagation is the Eisenbud-Wigner-Smith (EWS) time-delay \cite{Eisenbud:1948,Wigner:1955:PR:145,Smith:1960:PR:349}, which the photoelectron accumulates after its release into the continuum due to the interaction with the long-range Coulomb potential. The importance of the EWS time-delay in the photoemission from atoms, molecules and solids is highlighted in numerous review articles (see e.g. \cite{deCarvalho:2002:PR:83,Dahlstroem:2012:JPB:183001,Maquet:2014:JPB:204004,Pazourek:2015:RMP:765,Texier:2016:PE:16,Deshmukh:2021:TEPJST:1,Borrego-Varillas:2022:RPP:066401,Kheifets:2023:JPBAMOP:022001,Fetic:2024:AP:169666,Eckart:2024:JPBAMOP:202001}). An equally prominent example is the tunneling time-delay associated with the electron transition that occurs in the strong-field ionization regime and which is closely related to the generation of attosecond laser pulses by high-harmonic generation \cite{Chang:2016:JOSAB:1081}. It is attributed to the time which the electron spends in the classically forbidden region of the potential before its release into the continuum. For an overview on this topic, we refer the reader to the review articles \cite{Pfeiffer:2013:CP:84,Landsman:2015:PR:1,Hofmann:2019:JMO:1052,Sokolovski:2018:CP:1,Rost:2019:NP:439}. Recently, another time-delay associated with the electron transition in the multiphoton ionization (MPI) regime, has been discovered \cite{Su:2014:PRL:263002} and has subsequently attracted considerable attention both theoretically \cite{Pazourek:2015:JPBAMOP:061002,Argenti:2017:PRA:043426,Goldsmith:2018:JPBAMOP:155602} and experimentally \cite{Gong:2017:PRL:143203,Ge:2018:PRA:013409,Li:2024:PRA:013103}. This delay, referred to as absorption time-delay, occurs in the resonance-enhanced multiphoton ionization (REMPI). Compared to non-resonant MPI, which is considered to be quasi instantaneous \cite{Su:2014:PRL:263002}, in REMPI the photoemission is delayed due to the transient dynamics of the electron in the resonant intermediate states of the binding potential. The acquired time-delay was shown to be substantial and to scale linearly with the duration of the ionizing pulse \cite{Su:2014:PRL:263002}. This time-delay in REMPI induced by the intermediate resonances, which we refer to in the following as \textit{resonant ionization delay}, is the focus of the present study. \\
To date, photoionization time-delays have been studied mostly using either attosecond streaking \cite{Hentschel:2001:Nature:509,Drescher:2001:Science:1923,Sansone:2006:Science:443,Schultze:2010:Science:1685}, including the related angular streaking or `attoclock' technique \cite{Eckle:2008:NP:565,Landsman:2015:PR:1,Hofmann:2019:JMO:1052,SatyaSainadh:2020:JPP:042002}, or RABBITT (`reconstruction of attosecond beating by interference of two-photon transitions') \cite{Veniard:1996:PRA:721,Paul:2001:Science:1689,Varju:2005:L:888,Kluender:2011:PRL:143002}.  
While in streaking, the time-delay information is extracted directly from the temporal (or angular) shift of the streaking trace,  
in RABBITT the time-delay is derived from the spectral phase information encoded in the measured interferogram. The connection between the two approaches is provided by the group delay of the electron wave packet $\psi_e(t)$, which is related to the phase of its spectrum $\tilde{\psi}_e(\varepsilon)$ \cite{Dahlstroem:2012:JPB:183001,Maquet:2014:JPB:204004} according to
\begin{equation}
	\tau(\varepsilon)=-\hbar\frac{\mathrm{d}}{\mathrm{d}\varepsilon}\arg[\tilde{\psi}_e(\varepsilon)].
\end{equation}
This relation is the basis for the accurate determination of time-delays smaller than the duration of the ionizing laser pulse. \\
In this paper, we present a new approach for the determination of photoionization time-delays by the phase-sensitive detection of photoelectron wave packets. Our approach is based on the measurement of photoelectron vortices \cite{NgokoDjiokap:2015:PRL:113004,Pengel:2017:PRL:053003} created by REMPI. Specifically, we investigate the resonant ionization delay in the $(1+2)$ REMPI of potassium atoms via the $4p$-state, induced by polarization-shaped femtosecond (fs) laser pulses. In general, photoelectron vortices are created by ionization with a sequence of two time-delayed counter-rotating circularly polarized (CRCP) broadband laser pulses \cite{NgokoDjiokap:2015:PRL:113004,Pengel:2017:PRA:043426}. The spectral interference of the partial wave packets released by each pulse results in a spiral-shaped photoelectron momentum distribution (PMD), which corresponds to an angle-resolved Ramsey-type interferogram \cite{Ramsey:1950:PR:695,NgokoDjiokap:2015:PRL:113004}. The number of vortex arms, i.e., the rotational symmetry of the vortex, is determined by the angular momenta of the interfering partial wave packets. The slope of the vortex arms is determined by the relative time-delay between the partial wave packets and indicates the timing of the photoelectron wave packets with interferometric precision.\\
In REMPI involving a single resonance, the two partial wave packets created by ground state ionization via the resonant intermediate state experience the additional resonant ionization delay. Since the resonance-induced time-delay is the same for both partial wave packets, the pulse-to-pulse delay is encoded in the interferogram rather than the resonance-induced time-delay itself. Therefore, measuring the resonant ionization delay requires an independent reference that is not affected by the resonant excitation delay. This reference is provided by a third partial wave packet created via excitation of the resonant state by the first pulse and ionization from the resonant state by the second pulse. The reference wave packet is created quasi instantaneously by non-resonant ionization of the resonant state with the second pulse, i.e., without the resonant ionization delay. Pairwise interference with each of the other partial wave packets gives rise to two additional photoelectron vortices whose slopes encode the resonant ionization delay. To disentangle the three vortices, all of which are superimposed in the measured PMD, we make use of their symmetry properties. Because the reference wave packet has a different angular momentum, the three vortices are distinguished by their rotational symmetry. Building on our recent work on the measurement of spectral phases of shaped photoelectron vortices from MPI with chirped CRCP pulses \cite{Koehnke:2024b}, we decompose the PMD into its rotational symmetry components using 3D Fourier analysis. In this way, we reconstruct the different vortices and retrieve the resonant ionization delay from the slope of their vortex arms.
\begin{figure*}
	\includegraphics{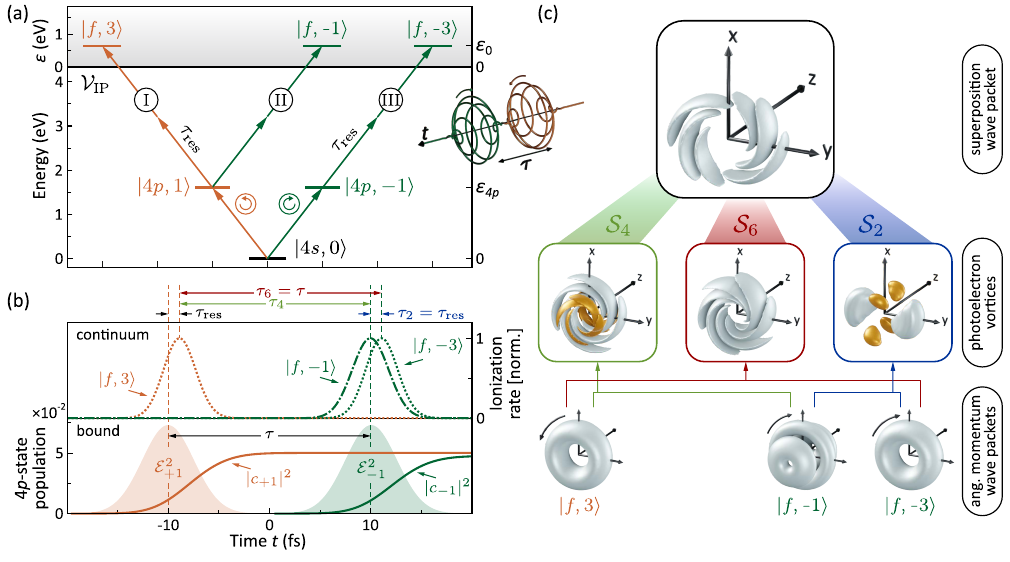}%
	\caption{Method for the determination of ionization time-delays in (1+2) REMPI using photoelectron vortices. (a) Excitation scheme for the interaction of potassium atoms with a time-delayed CRCP pulse sequence. Photoelectron partial wave packets created via the ionization pathways (I) and (III) accumulate the same resonant ionization delay $\tau_\mathrm{res}$. In contrast, the photoelectron partial wave packet created via the sequential ionization pathway (II) is not delayed, serving as a reference. (b) Population dynamics of the resonantly excited states $|4p,\pm1\rangle$ (lower frame; solid lines) and time-resolved ionization rate $\mathcal{R}(t)$ to the different continua (upper frame; dotted and dashed-dotted lines). (c) Densities of the created angular momentum partial wave packets (lower row). Pairwise interference of the partial wave packets gives rise to three photoelectron vortices with distinct rotational symmetry and individual slope of their spiral arms, which encode the relative time-delay of the interfering partial wave packets (middle row). All vortices are superimposed in the total PMD observed in the experiment (top row).\label{fig1}}
\end{figure*}

\section{Physical system\label{sec:PhysicalSystem}}
We study the resonant ionization delay in the perturbative (1+2) REMPI of potassium atoms, using photoelectron vortices \cite{Pengel:2017:PRL:053003} created by a sequence of CRCP white-light supercontinuum pulses \cite{Pengel:2017:PRA:043426}. The corresponding ionization scheme is depicted in Fig.~\ref{fig1}(a). For simplicity, we consider only $\Delta\ell=+1$ transitions, motivated by Fano's propensity rule \cite{Fano:1985:PRA:617}. The laser electric field of the CRCP sequence is described as
\begin{equation}
	\boldsymbol{E}(t) = \left[\mathcal{E}_{+1}(t)\boldsymbol{e}_{+1}+\mathcal{E}_{-1}(t)\boldsymbol{e}_{-1}\right]e^{-i\omega_0t},
\end{equation}
where $\mathcal{E}_\mathrm{\pm1}(t)=\mathcal{E}(t\pm\frac{\tau}{2})$ are the shifted envelopes of the left- (LCP) and right-handed circularly polarized (RCP) pulse components, $\tau$ is the pulse-to-pulse delay, $\omega_0$ is the laser central frequency and $\boldsymbol{e}_{\pm1}$ are the Jones vectors for LCP and RCP light, respectively. Without loss of generality, we consider a first LCP pulse followed by a second RCP pulse. Following the dipole selection rules, the first pulse drives (1+2) REMPI via the $|4p,1\rangle$-state and creates an $|f,3\rangle$-type photoelectron partial wave packet. This ionization pathway is labeled (I). Due to the resonance, after the pulse, some population is stored in the $|4p,1\rangle$ state. Therefore, the time-delayed second pulse ionizes the atom along two pathways. In analogy to pathway (I), an $|f,-3\rangle$-type photoelectron partial wave packet is created by (1+2) REMPI via the $|4p,-1\rangle$-state [ionization pathway (III)].
In addition, the second pulse ionizes the atom from the excited $|4p,1\rangle$ state creating an $|f,-1\rangle$-type photoelectron partial wave packet along pathway (II), which serves as the reference for measuring of the ionization time delay.\\
The energy-dependent amplitudes of the photoelectron partial wave packets are determined using time-dependent perturbation theory, analogous to the treatment in \cite{Eickhoff:2022:PRA:053113}. To this end, we consider a three-state atom consisting of the ground state $|4s,0\rangle$ and the degenerate bound states $|4p,\pm1\rangle$. 
The amplitudes of the $|f,\pm3\rangle$-type partial wave packets, created along pathways (I) and (III), are given by second-order perturbation theory as \cite{Meier:1994:PRL:3207}
\begin{equation}
	a_{\pm3}(\varepsilon)=-\frac{\mu^{(2)}}{\hbar^2}\int\limits_{-\infty}^\infty c_{\pm1}(t)\mathcal{E}^2_{\pm1}(t)e^{-i(\varepsilon-\varepsilon_0)\frac{t}{\hbar}}\mathrm{d}t.
	\label{eq:PEamplitudes_3}
\end{equation}
Herein, $\mu^{(2)}$ denotes the effective radial two-photon transition dipole moment of the $4p$-states to the $f$-type ionization continua, $\varepsilon$ is the photoelectron kinetic energy and $\varepsilon_0=3\hbar\omega_0-\mathcal{V}_\mathrm{IP}$ describes the center of the photoelectron energy distribution, with $\mathcal{V}_\mathrm{IP}$ being the ionization potential. The functions $c_{\pm1}(t)$ are the time-dependent population amplitudes of the $4p$-states in the interaction picture (cf. Eq.~\eqref{eq:4p_amplitudes}). Equation~\eqref{eq:PEamplitudes_3} describes the (1+2) REMPI process as an interplay between the excitation of the $4p$-states and the simultaneous two-photon ionization from the $4p$-states, mapping the $4p$-population dynamics into the continuum. The product of the $4p$-amplitudes $c_{\pm1}(t)$ and the second harmonic envelopes $\mathcal{E}^2_{\pm1}(t)$ in the integrand of Eq.~\eqref{eq:PEamplitudes_3} describes the ionization rate as
\begin{equation}\label{eq:ionization_rate}
	\mathcal{R}(t)=\left|c_{\pm1}(t)\mathcal{E}^2_{\pm1}(t)\right|^2.
\end{equation}
Physically, we interpret this product as a gating of the two-photon ionization field $\mathcal{E}^2_{\pm1}(t)$ by the excitation dynamics $c_{\pm1}(t)$ in the $4p$-states. In the weak-field regime, the $4p$-amplitudes are given by first-order perturbation theory as 
\begin{equation}
	c_{\pm1}(t)=-\frac{\mu_0}{i\hbar}\int\limits_{-\infty}^t\mathcal{E}_\mathrm{\pm1}(t')e^{-i\delta t'}\mathrm{d}t',
	\label{eq:4p_amplitudes}
\end{equation}
where $\mu_0$ is the $4s\rightarrow4p$ transition dipole moment and $\delta=\omega_0-\omega_{4p}$ is the detuning of the laser pulse from the $4s\rightarrow4p$ transition. For clarity, here we consider resonant excitation, i.e. $\delta=0$, and address the influence of the detuning later in Sec.~\ref{sec:Results} (see also Fig.~\ref{fig4}). 
In the resonant case, the populations $|c_{\pm1}(t)|^2$ build up sigmoidally during the respective pulse, as shown in the lower panel of Fig.~\ref{fig1}(b) for a sequence of two Gaussian-shaped CRCP pulses with a full width at half maximum (FWHM) of the intensity of $\Delta t=\SI{6}{fs}$ and a pulse-to-pulse delay of $\tau=\SI{20}{fs}$.
Due to the asymmetric gating of the two-photon ionization process, the peak of the ionization rate $\mathcal{R}(t)$, shown as dotted curves in the upper panel of Fig.~\ref{fig1}(b), is shifted from the maximum of the pulse envelopes to later times. This temporal shift causes the delayed emission of the respective photoelectron partial wave packets. Since Eq.~\eqref{eq:PEamplitudes_3} has the form of a Fourier transform, the temporal shift of the ionization rate manifests in the spectral phase imprinted in the amplitudes $a_{\pm3}(\varepsilon)$. In the following, the resonant ionization delay between the birth of the photoelectron partial wave packets and the center of the ionizing pulse is denoted as $\tau_\mathrm{res}$ as indicated in Fig.~\ref{fig1}(b). As elucidated in the introduction, $\tau_\mathrm{res}$ is extracted from the photoelectron amplitudes as
\begin{align}
	\tau_\mathrm{res}=-\hbar\frac{\mathrm{d}}{\mathrm{d}\varepsilon}\arg\left[a_{\pm3}(\varepsilon)\right]\pm\frac{\tau}{2}.
	\label{eq:tau_res}
\end{align}
For a Gaussian-shaped envelope $\mathcal{E}(t)=\mathcal{E}_0\,e^{-\ln(4)\left(\frac{t}{\Delta t}\right)^2}$ with an intensity FWHM of $\Delta t$, Eq.~\eqref{eq:tau_res} can be evaluated analytically (see Appendix \ref{sec:App_AnaDelayGauss}) yielding
\begin{equation}
	\tau_\mathrm{res}=\frac{\Delta t}{2\sqrt{\pi\ln(8)}}\approx 0.2\,\Delta t.
	\label{pyssys:Gaussian}
\end{equation}
Equation~\eqref{pyssys:Gaussian} shows that the resonant ionization delay in the (1+2) REMPI scenario depends linearly on the pulse duration $\Delta t$, in agreement with the results reported in \cite{Su:2014:PRL:263002,Goldsmith:2018:JPBAMOP:155602}. Similar analytical results, obtained for ($\REMPIM$+$\REMPIN$) REMPI with a Gaussian-shaped envelope and for other pulse shapes are summarized in the Appendix in the sections \ref{sec:App_AnaDelayGaussNM} and \ref{sec:App_AnaDelayShape}. However, since the resonant ionization delay is accumulated in both pathway (I) and pathway (III), the interference between the corresponding partial wave packets $\psi_{3,3}$ and $\psi_{3,-3}$ is not sensitive to $\tau_\mathrm{res}$. In fact, the analysis of the resulting $6$-arm vortex (interferogram) yields the pulse-to-pulse delay $\tau$. \\
Key to the retrieval of $\tau_\mathrm{res}$ is the sequential ionization pathway (II) which is not gated by the $4p$-population dynamics, provided that the ionizing second pulse is temporally well-separated from the exciting first pulse. In this case, the direct two-photon ionization maps the final population of the $|4p,1\rangle$ state. The final amplitude of the $|4p,1\rangle$ state, defined by Eq.~\eqref{eq:4p_amplitudes} as $t\rightarrow \infty$ reads
\begin{equation}
	c_{1}^\infty=c_{1}(\infty) =-\frac{\mu_0}{i\hbar}\tilde{\mathcal{E}}(0),
\end{equation}
where $\tilde{\mathcal{E}}(\omega)=\mathcal{F}[\mathcal{E}(t)]$ is the spectrum of the pulse envelope.
The amplitude of the created $|f,-1\rangle$-type photoelectron partial wave packet is therefore given by
\begin{equation}
	a_{-1}(\varepsilon)=-\frac{\mu^{(2)}}{\sqrt{15}\,\hbar^2}\int\limits_{-\infty}^\infty c_{1}^\infty\mathcal{E}^2_{-1}(t)e^{-i(\varepsilon-\varepsilon_0)\frac{t}{\hbar}}\mathrm{d}t.
	\label{eq:PEamplitude_1}
\end{equation}
The factor $1/\sqrt{15}$ originates from the angular part of the two-photon transition dipole moment. According to Eq.~\eqref{eq:PEamplitude_1}, the ionization rate instantaneously follows the second harmonic envelope $\mathcal{E}^2_{-1}(t)$, scaled by $c_{1}^\infty$, which is seen by inspecting the dashed-dotted curve in the top frame of Fig.~\ref{fig1}(b). 
Due to its instantaneous creation, the partial wave packet $\psi_{3,-1}$ generated along pathway (II) is the temporal reference. Its interference with the partial wave packets $\psi_{3,\pm3}$ gives rise to a $4$-arm vortex and a $2$-arm vortex, respectively, in which the resonant ionization delay $\tau_\mathrm{res}$ is encoded. \\
Finally, we briefly discuss our procedure for retrieving the resonant ionization delay from the measured PMD.
A more detailed description is given in \cite{Koehnke:2024b}. The procedure is illustrated in Fig.~\ref{fig1}(c). The PMD is proportional to the 3D probability density of the total photoelectron wave packet, shown in the top row of Fig.~\ref{fig1}(c), resulting from the superposition of the three partial wave packets shown in the bottom row of Fig.~\ref{fig1}(c).
The density is written as
\begin{align}
	\mathcal{P}(\varepsilon,&\vartheta,\phi)=|\psi_{3,3}+\psi_{3,1}+\psi_{3,-3}|^2\notag\\
	&=\mathcal{S}_0(\varepsilon,\vartheta)+\mathcal{S}_2(\varepsilon,\vartheta,\phi)+\mathcal{S}_4(\varepsilon,\vartheta,\phi)+\mathcal{S}_6(\varepsilon,\vartheta,\phi).
	\label{eq:density}
\end{align}
The first term $\mathcal{S}_0(\varepsilon,\vartheta)$ is an azimuthally isotropic background, while the terms $\mathcal{S}_j(\varepsilon,\vartheta,\phi)\propto\cos(j\,\phi+\frac{\varepsilon}{\hbar}\tau_j)$ are photoelectron vortices of $j$-fold azimuthal symmetry arising from the pairwise interference of the partial wave packets, as shown in the middle row of Fig.~\ref{fig1}(c). The $\tau_j$ are the relative time-delays between the partial wave packets, given explicitly by
\begin{align}
	\tau_6 &= \tau,\notag\\
	\tau_4 &= \tau-\tau_\mathrm{res},\notag\\
	\tau_2 &= \tau_\mathrm{res},\label{eq:EffectiveDelays}
\end{align}
as illustrated in Fig.~\ref{fig1}(b).
To retrieve the relative time-delays from the vortices, the PMD is decomposed into its symmetry components by 3D Fourier analysis in azimuthal direction, as recently described in \cite{Koehnke:2024b}. Then, according to Eq.~\eqref{eq:EffectiveDelays}, the resonant ionization delay is obtained as the difference between the time-delays encoded in the $c_6$- and $c_4$-symmetric vortices, $\tau_\mathrm{res}=\tau_6-\tau_4$, and additionally encoded directly in the $c_2$-symmetric vortex, which allows us to validate the two results against one another.

\section{Experimental Setup\label{sec:Setup}}
We briefly describe our experimental setup and procedure in this section. A detailed description is given in \cite{Pengel:2017:PRA:043426,Koehnke:2024b}. The white-light supercontinuum is generated by seeding a neon-filled hollow-core fiber (absolute pressure ~\SI{2}{\bar}) with \SI{20}{fs} infrared pulses from a chirped pulse amplification fs-laser system (\textsc{FEMTOLASERS RAINBOW 500}, \textsc{CEP4 Module}, \textsc{FEMTOPOWER HR CEP}) with a central wave length of \SI{790}{\nano\meter}, a repetition rate of \SI{3}{\kilo\hertz} and a pulse energy of \SI{1.0}{\milli\joule}. Using a home-built white-light polarization pulse shaper in $4f$-geometry \cite{Kerbstadt:2017:OE:12518}, we apply linear spectral phases to two orthogonal polarization components of the input pulse to generate an orthogonal linearly polarized (OLP) pulse sequence. A superachromatic $\lambda/4$ waveplate converts the OLP sequence into a CRCP sequence. The CRCP sequence is focused into the interaction region of a velocity map imaging (VMI) spectrometer \cite{Eppink:1997:RSI:3477}, filled with potassium vapor from a dispenser source, using a spherical mirror ($f=\SI{250}{\milli\meter}$). The peak intensity in the interaction region is estimated to be about \SI{5e12}{\watt\per\centi\meter^2}, which corresponds to perturbative interaction conditions \cite{Karule:1990:265}. The generated PMD is imaged onto a two-dimensional (2D) position-sensitive detector (German Image Detector Systems MCP-77-2-60-P43-CF160-HR) and the 2D projections are recorded using a coupled charge device (CCD) camera (Lumenera LW165M). For the reconstruction of the 3D PMD, we use the photoelectron tomography technique described in \cite{Wollenhaupt:2009:APB:647}. \\
The reconstructed PMD is decomposed into its azimuthal symmetry components by 3D Fourier analysis \cite{Koehnke:2024b}. The procedure yields the energy-dependent amplitudes and phases of the $c_2$-, $c_4$- and $c_6$-symmetric photoelectron vortices. The linear regression of the spectral phases as a function of the photoelectron kinetic energy finally yields the time-delays $\tau_2$, $\tau_4$ and $\tau_6$, from which we derive the resonant ionization delay $\tau_\mathrm{res}$ according to Eq.~\eqref{eq:EffectiveDelays}.
\begin{figure}[t]
	\includegraphics[width=\linewidth]{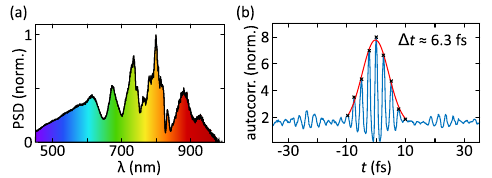}
	\caption{Characterization of the white-light supercontinuum pulses used in the experiment. (a) Measured power spectral density (PSD). (b) Shaper-based second-order autocorrelation trace. Considering a Gaussian-shaped pulse, we estimate an intensity FWHM pulse duration of $\Delta t=\SI{6.3(10)}{\femto\second}$ by fitting a Gaussian (red line) to the local maxima (crosses) in the time interval $t\in[\SI{-10}{\femto\second}, \SI{10}{\femto\second}]$.  \label{fig2}}
\end{figure}
\begin{figure*}[t]
	\includegraphics[width=\linewidth]{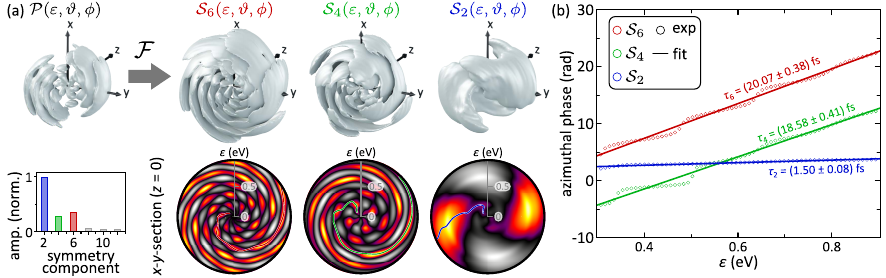}
	\caption{Decomposition of the measured and tomographically reconstructed 3D PMD into its major symmetry components $c_6$, $c_4$ and $c_2$. (a) 3D Fourier analysis of the PMD yields the amplitudes of the vortices $\mathcal{S}_j(\varepsilon,\vartheta,\phi)$ represented in the upper row. The lower row shows central $x$-$y$-sections ($z=0$) through the vortices. The colored solid lines highlight the energy-dependent rotation angles of the spiral arms. (b) Extracted energy-dependent spectral phases (open circles) of the vortices together with linear fits (solid lines) in the interval $\varepsilon\in[0.3,0.9]\mathrm{eV}$. The effective time-delays $\tau_j$ retrieved from the fits are indicated.\label{fig3}}
\end{figure*}

\section{Results\label{sec:Results}}
For our investigation of the resonant ionization time-delay in (1+2) REMPI of potassium atoms, we tuned the central frequency of the white-light supercontinuum to the $4s$-$4p$-resonance at $\omega_0=\SI{2.45}{\radian\per\femto\second}$. A measured spectrum of the white-light supercontinuum is shown in Fig.~\ref{fig2}(a).

\begin{figure}[t]
	\includegraphics[width=\linewidth]{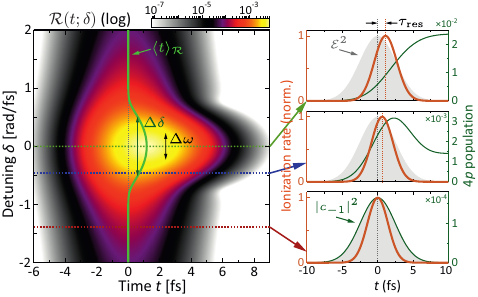}
	\caption{Numerical simulation of the ionization rate $\mathcal{R}(t;\delta)$ from Eq.~\eqref{eq:ionization_rate} as a function of time and the detuning from resonance, for a Gaussian-shaped single pulse of duration $\Delta t=\SI{6}{\femto\second}$. The blue solid line indicates the first moment $\langle t\rangle_\mathcal{R}(\delta)$. For resonant excitation, $\delta=0$, the ionization rate is significantly enhanced. The maximum is shifted in time by approximately $\tau_\mathrm{res}$ due to the asymmetric gating of the ionizing field $\mathcal{E}^2(t)$ by the sigmoidal $4p$-dynamics (upper right inset). For non-resonant excitation, $\delta\gg\Delta\omega=4\ln(2)/\Delta t$, the $4p$-population follows the the pulse intensity envelope (lower right inset) and the ionization rate is not delayed in time.\label{fig4}}
\end{figure}

The spectral bandwidth corresponds to a Fourier-limited pulse duration of $\Delta t=\SI{6}{\femto\second}$. We verified this estimate by a second-order autocorrelation measurement, using the pulse-shaper to generate two replica of the input pulse with a variable time-delay and focusing the generated pulse sequences into a $\beta$-barium borate crystal for second harmonic generation \cite{Koehler:2011:OE:11638,Kerbstadt:2017:OE:12518}. From the autocorrelation trace, recorded using a spectrometer and shown in Fig.~\ref{fig2}(b), we retrieve a pulse duration of about $\Delta t=\SI{6.3}{fs}$. In the actual experiment, the pulse-to-pulse delay was set to $\tau=\SI{20}{\femto\second}$ to ensure that the pulses in the CRCP sequence are well-separated in time. The experimental results are presented in Fig.~\ref{fig3}. The top row of Fig.~\ref{fig3}(a) shows the tomographically reconstructed and energy-calibrated 3D PMD $\mathcal{P}(\varepsilon,\vartheta,\phi)$ (left). The shape of the PMD is predominantly a $6$-arm spiral, as expected for the three-photon ionization scenario \cite{Pengel:2017:PRL:053003,Pengel:2017:PRA:043426}. However, the presence of additional symmetry components is already visible from the non-homogeneous intensity of the vortex arms. From the apparent $c_2$-symmetric modulation of the PMD, we deduce that the $c_2$-symmetry component has a large contribution. 3D Fourier analysis of the PMD in the azimuthal direction yields the amplitudes of the symmetry components, shown in the frame below the PMD in Fig.~\ref{fig3}(a). The distribution shows that the PMD is mainly composed of the $c_6$-, $c_4$- and $c_2$-symmetry components. Minor contributions from the $c_8$- and higher components are attributed to experimental imperfections, such as noise of the measured VMI projections. The amplitudes of the odd-numbered symmetry components were also found to be very small as suggested by parity considerations: PMDs created by single-color MPI are expected to be left-right and forward-backward symmetric with respect to the laser propagation direction \cite{Eickhoff:2021:FP:444}. Therefore, for the analysis, we symmetrized the measured projections along the $y$- and the $z$-direction and utilized the redundancy of the data to improve the measurement statistics. \\
The three major symmetry components $\mathcal{S}_j(\varepsilon,\vartheta,\phi)$ ($j=2,4,6$) obtained by 3D Fourier decomposition
are visualized on the right-hand side of Fig.~\ref{fig3}(a). All three components have a spiral shape and encode information about the relative time-delays between the interfering photoelectron partial wave packets. We analyze the azimuthal phase of the three vortices as a function of the photoelectron energy $\varepsilon$ to extract these time-delays. Our 3D Fourier analysis method retrieves the spectral phase of each symmetry component as a 2D function of the polar angle $\vartheta$ and the energy $\varepsilon$. To obtain a clear picture of their energy dependence, we integrate these phase functions over a small polar angle segment $\vartheta\in[\SI{80}{\degree},\SI{100}{\degree}]$. The segment is sufficiently small to avoid blurring of the $c_2$- and $c_4$-spiral patterns due the $\pi$-jump in the polar part of the reference wave packet $\psi_{3,-1}$ \cite{Pengel:2017:PRA:043426,Koehnke:2022:JPBAMOP:184003}. The latter is determined by the spherical harmonic $Y_{3,-1}(\vartheta,\phi)$ with its angular nodes around $\vartheta=\SI{63.4}{\degree}$ and $\SI{116.6}{\degree}$.  
When divided by the respective vortex symmetry, the energy-dependent spectral phases give the azimuthal rotation angle of the spiral arms as a function of $\varepsilon$ \cite{Kerbstadt:2019:NC:658}. This is illustrated in the bottom row of Fig.~\ref{fig3} by the colored curves accurately tracing the spiral arms in the $x$-$y$-sections through the corresponding vortices.
The extracted energy-dependent phase functions are depicted as circles in Fig.~\ref{fig3}(b). In the energy window $\varepsilon\in[\SI{0.3}{\electronvolt},\SI{0.9}{\electronvolt}]$, where the vortices are centered, the phases are indeed linear -- as expected from the discussion in Sec.~\ref{sec:PhysicalSystem}. The slopes of the linear functions encode the relative time-delays between the interfering wave packets. By applying linear fits of the form
\begin{equation}
\varphi_j(\varepsilon)=\frac{\tau_j}{\hbar}\varepsilon+\varphi_0,
\end{equation}
with $j=2,4,6$ to the phase functions (solid lines in Fig.~\ref{fig3}(b)), we retrieve the respective relative time-delays as
\begin{align}
	\tau_6^{\mathrm{exp}} & = \SI{20.07(38)}{\femto\second},\notag\\
	\tau_4^{\mathrm{exp}} & = \SI{18.58(41)}{\femto\second},\notag\\
	\tau_2^{\mathrm{exp}} & = \SI{1.50(8)}{\femto\second}.
	\label{eq:exptaus}
\end{align}
The time-delay $\tau_6^\mathrm{exp}$ of the $c_6$-symmetric vortex agrees very well with the pulse-to-pulse delay $\tau$ introduced experimentally by the pulse shaper and being in accordance with the theoretical value predicted by our analytical model in Eq.~\eqref{eq:EffectiveDelays}. According to the model, the time-delay $\tau_4^\mathrm{exp}$ encoded in the $c_4$-symmetric vortex, is given by the difference between the pulse-to-pulse delay and the resonant ionization delay $\tau_\mathrm{res}$. The difference of $\tau_6^\mathrm{exp}$and $\tau_4^\mathrm{exp}$ is in excellent agreement with the retrieved time-delay $\tau_2^\mathrm{exp}=\SI{1.50(8)}{\femto\second}$ of the $c_2$-symmetric vortex in Eq.~\eqref{eq:exptaus}, as also predicted by our theoretical model. We estimate the resonant ionization time-delay by the mean value of $\tau_6-\tau_4$ and $\tau_2$, resulting in $\tau_\mathrm{res}^\mathrm{exp}=\SI{1.50(44)}{\femto\second}$. \\
Equation~\eqref{pyssys:Gaussian} suggests a resonant ionization delay of $\tau_\mathrm{res}^{\mathrm{theo}}=\SI{1.17}{\femto\second}$ for (1+2) REMPI by a Gaussian-shaped pulse with an FWHM duration of $\Delta t=\SI{6}{\femto\second}$. Although the theoretical value is in agreement with the experimental result within its error range, we briefly discuss possible reasons for the deviations between $\tau_\mathrm{res}^\mathrm{exp}$ and $\tau_\mathrm{res}^\mathrm{theo}$. First, the white-light spectrum used in the experiment is not Gaussian-shaped. As discussed in Appendix~\ref{sec:App_AnaDelay}, the value of the resonant ionization delay depends sensitively on the pulse shape. Considering different pulse shapes of the same FWHM duration $\Delta t$, we obtain values for the delay ranging from $\tau_\mathrm{res}^{\mathrm{theo}}=\SI{1.00}{\femto\second}$, for a rectangular pulse, to $\SI{1.48}{\femto\second}$ for a Lorentzian pulse. Second, our theoretical model does not account for near-resonant intermediate states in the two-photon ionization from the $4p$-state. Potential candidates in the potassium atom are the $3d$-, $4d$- and $5d$-states, all of which lie within the bandwidth of the second harmonic spectrum of the white-light pulses. These states are either weakly coupled ($4d$) or significantly detuned ($3d$, $5d$). Nevertheless, they could play a role in the excitation dynamics and cause additional time-delays in the photoemission \cite{Goldsmith:2018:JPBAMOP:155602}. \\
Finally, we discuss the transition from the (1+2) REMPI scenario considered so far to non-resonant three-photon ionization, where photoemission is assumed to occur quasi-instantaneously \cite{Su:2014:PRL:263002}. Motivated by the good agreement between our theoretical model and the experimental results, we use our numerical simulations to investigate the dependence of the resonant ionization delay $\tau_\mathrm{res}$ on the detuning $\delta$ (cf. Eq.~\eqref{eq:4p_amplitudes}) of the laser pulse from the $4p$-resonance. For this purpose, we considered a Gaussian-shaped single pulse centered at $t=0$, with a duration of $\Delta t=\SI{6}{fs}$ and variable central frequency $\omega_0$, and calculated the time-dependent ionization rate $\mathcal{R}(t;\delta)$ (see Eq.~\ref{eq:ionization_rate}) as a function of the detuning $\delta$. The resulting 2D spectrogram is shown in Fig.~\ref{fig4} in logarithmic representation. The 2D-trace exhibits a distinct global maximum centered on resonance, i.e. $\delta=0$ and shifted in time by approximately (not exactly) the resonant ionization delay $\tau_\mathrm{res}$.
For blue- ($\delta>0$) and red-detuned ($\delta<0$) excitation, the ionization rate decays symmetrically towards zero. To highlight the detuning-dependence of photoionization time-delay in the 2D-trace, we calculated the first moment of the ionization rate 
\begin{equation}
	\langle t\rangle_\mathcal{R}(\delta)=\frac{\intop_{-\infty}^\infty t\,\mathcal{R}(t;\delta)dt}{\intop_{-\infty}^\infty \mathcal{R}(t;\delta)dt}.
\end{equation}
The result is plotted as light-green solid line over the 2D-trace. The FWHM of the bell-shaped curve, indicated by the green vertical double-headed arrow, is determined numerically to be $\Delta\delta=\SI{1.03}{rad/fs}$. In comparison, the spectral bandwidth of the laser pulse is given by $\Delta\omega=2\ln(4)/\Delta t=\SI{0.46}{rad/fs}$ (black double-headed arrow), i.e., significantly smaller than the width $\Delta\delta$ of the $\langle t\rangle_\mathcal{R}(\delta)$ curve.
Thus, even for detunings in the order of the pulse bandwidth $\Delta\omega$, the $4p$-resonance induces a considerable ionization time-delay. \\
The maximum time-delay is obtained for $\delta=0$ (horizontal green dotted line). The upper inset to Fig.~\ref{fig4} shows the corresponding time-dependent population of the $4p$-state, the second harmonic envelope $\mathcal{E}^2(t)$ governing the two-photon ionization and the resulting ionization rate. Here, the second harmonic is gated asymmetrically by the sigmoidal build-up of the $4p$-population, similar to the discussion of Fig.~\ref{fig1}. In contrast, in the spectral wings of the trace, e.g. for $\delta=-\Delta\omega$ (blue dotted line), the population of the $4p$-state evolves non-monotonously, as shown in the middle inset to Fig.~\ref{fig4}. The population transfer peaks shortly after the maximum of the pulse and, subsequently, decreases again to reach a lower final population than in the resonant case. The smaller temporal shift in the ionization rate results from the gating of the second harmonic by the characteristic hump-like transient. Eventually, for detunings that significantly exceed the spectral bandwidth of the pulse, e.g. for $\delta=-3\Delta\omega$ (red dotted line), the transient $4p$-population follows the pulse intensity $|\mathcal{E}(t)|^2$ such that the second harmonic is gated symmetrically in time. In this case, the maximum of the ionization rate coincides with the maximum of the pulse at $t=0$, i.e. the photoionization is not delayed, as shown in the lower inset to Fig.~\ref{fig4}.
\section{Summary and Conclusion\label{Conclusion}}
In this paper, we have studied the resonant ionization time-delay in the (1+2) REMPI of potassium atoms, which arises due to the transient dynamics of the electron in the resonant $4p$-state. Our results show that REMPI via a resonant intermediate state is associated with a significant photoemission time-delay, in other words \emph{resonance takes time}. We used a new technique based on the phase-sensitive detection of photoelectron vortices to measure the resonant ionization delay.
These spiral-shaped photoelectron wave packets, produced by Ramsey-type interference of two photoelectron partial wave packets with different angular momenta, are highly sensitive to the timing of the photoemission events. The timing information contained in the spectral phase is encoded in the slope of the vortex arms with interferometric precision. Key to measuring the resonant ionization delay is the superposition of the above-mentioned partial wave packets with an additional reference wave packet. The reference wave packet is provided by the non-resonant ionization from the already populated  resonant intermediate state. Pairwise interference of all partial wave packets gives rise to three vortices of different rotational symmetry, superimposed to produce the total PMD. In the experiment, we reconstruct the 3D PMD tomographically and decompose the reconstructed PMD into the different azimuthal symmetry components using 3D Fourier analysis. From the retrieved spectral phase of the vortices, we obtain a resonant ionization delay in the (1+2) REMPI scenario of $\tau^\mathrm{exp}_\mathrm{res}=\SI{1.50(44)}{\femto\second}$, corresponding to about $20\%$ of the duration of the white-light supercontinuum pulses. This finding is in good agreement with an analytical model based on the perturbative (1+2) REMPI of a three-state atom by a Gaussian-shaped CRCP pulse sequence. Resonant ionization delays for ($\REMPIM$+$\REMPIN$) REMPI are presented for various pulse envelope shapes.\\
Our results show that photoelectron vortices are a powerful tool for the determination of photoionization time-delays on the femto- to sub-femtosecond timescale. In the light of recent advances made in the generation of polarization-shaped attosecond pulses {\cite{Jimenez-Galan:2018:NC:850,Huang:2018:NP:349}, our method opens up new prospects for photoelectron chronoscopy \cite{Pazourek:2015:RMP:765} and the measurement of photoionization time-delays on the attosecond timescale.

\begin{acknowledgments}
	We gratefully acknowledge financially support from the collaborative research program ``Dynamics on the Nanoscale'' (DyNano) and the Wissenschaftsraum ``Elektronen-Licht-Kontrolle'' (elLiKo) funded by the the Nieders{\"a}chsische Ministerium f{\"u}r Wissenschaft und Kultur.
\end{acknowledgments}


\appendix
\section{Resonant ionization delays: analytical description\label{sec:App_AnaDelay}}
\begin{figure}[t]
	\includegraphics[width=\linewidth]{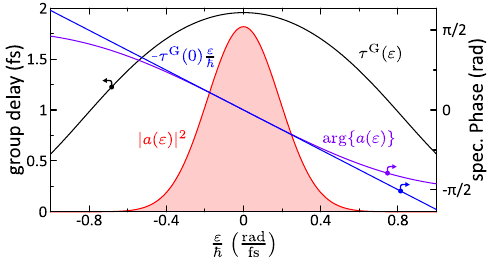}
	\caption{Spectral phase and ionization delay in (1+2) REMPI for a Gaussian shaped pulse with a pulse duration of $\Delta t=\SI{10}{\femto\second}$. The PSD of the photoelectron amplitude $a(\varepsilon)$ is depicted as red area in the background, while its phase is shown in violet. The corresponding linear spectral phase $-\tau^\mathrm{G}(0)\frac{\varepsilon}{\hbar}$ is given as a blue curve. The group delay $\tau^\mathrm{G}(\varepsilon)$ is shown as black solid line.\label{fig5}}
\end{figure}
In this section, we provide an analytical description for resonant ionization delays in various REMPI scenarios. At first, we consider (1+2)~REMPI with Gaussian-shaped pulses, motivated by the experiment. Next, the effects of the pulse shape and the number of photons used for excitation and ionization on the resonant ionization delay are investigated.

\subsection{(1+2) REMPI with Gaussian pulses} \label{sec:App_AnaDelayGauss}
Considering the photoelectron amplitudes $a(\varepsilon)$, we illustrate the calculation of the ionization time-delay for (1+2) REMPI with a Gaussian shaped laser pulse. A Gaussian electric field envelope with an intensity FWHM of $\Delta t$ reads
\begin{equation}
\mathcal{E}(t) = \mathcal{E}_0 \, e^{- \left(\beta_G \frac{t}{\Delta t} \right)^2},
\label{eq:AppGaussian}
\end{equation}
where $\beta_G = \sqrt{\ln(4)}$. Following the description in Sec.~\ref{sec:PhysicalSystem}, we apply first-order perturbation theory to derive the time-dependent amplitude of the $4p$-state
\begin{equation}
c_{4p}(t)=-\frac{\mu_0}{i\hbar}\int\limits_{-\infty}^t\mathcal{E}(t')\mathrm{d}t'.
\label{eq:AppPopulation4p}
\end{equation}
For the Gaussian pulse shape as defined in \eqref{eq:AppGaussian} we obtain
\begin{equation}
c_{4p}(t) = i \frac{\sqrt{\pi}}{2} \frac{\mathcal{E}_0 \, \mu_0}{\hbar}  \frac{\Delta t}{\beta_G} \left[1 + {\rm erf}\left(\frac{\beta_G}{\Delta t} \, t  \right) \right].
\end{equation}
Assuming perturbative non-resonant two-photon ionization of the $4p$-state and using $\varepsilon_0=0$ in Eq.~\eqref{eq:PEamplitudes_3} for simplicity 
\begin{equation}
a(\varepsilon)=-\frac{\mu^{(2)}}{\hbar^2}\int\limits_{-\infty}^\infty c_{4p}(t)\mathcal{E}^2(t)e^{-i \varepsilon \frac{t}{\hbar}}\mathrm{d}t,
\label{eq:AppGeneralPhotoelectronAmplitude}
\end{equation}
yields the photoelectron amplitudes for the Gaussian pulse
\begin{align}
a(\varepsilon) = &- \frac{i \pi}{2 \sqrt{2}} \mu_0 \mu^{(2)} 
\left( \frac{\mathcal{E}_0}{\hbar} \right)^3  
\left( \frac{\Delta t}{\beta_G} \right)^2 \notag \\
				& \times e^{- \frac{1}{8} \left( \frac{\Delta t \, \varepsilon}{ \beta_G \hbar}  \right)^2} \, \left[ 1 - \, {\rm erf} \left(  \frac{i}{2 \sqrt{6}} \frac{\Delta t \, \varepsilon}{ \beta_G \hbar}  \right) \right].
				\label{eq:AppGaussianPhotoelectronAmplitude}
\end{align}
Figure~\ref{fig5} shows the power spectral density (PSD) of the photoelectron amplitude as red solid line and its spectral phase in black. The group delay (purple) is derived from Eq.~\eqref{eq:AppGaussianPhotoelectronAmplitude} to give
\begin{align}
\tau^G(\varepsilon) &= - \hbar \frac{\mathrm{d}}{\mathrm{d}\varepsilon} \arg[a(\varepsilon)] \notag\\
&= \frac{1}{\sqrt{6 \pi} } \,
  \frac{\Delta t }{\beta_G } \;
   \frac{ e^{\frac{1}{24} \left( \frac{\Delta t \, \varepsilon}{ \beta_G \hbar}  \right)^2}  }{  1 - \, \left[ {\rm erf} \left(  \frac{i}{2 \sqrt{6}} \frac{\Delta t \, \varepsilon}{ \beta_G \hbar}  \right)  \right]^2  }.
 \label{eq:AppGaussianGroupDelay}
\end{align}
The linear spectral phase at the laser central frequency  $\varepsilon=0$ (blue line in Figure~\ref{fig5}) results in the resonant ionization delay of
\begin{equation}
\tau^G(0) = \frac{1}{\sqrt{6 \pi}}  \frac{\Delta t }{\beta_G} \approx 0.196 \Delta t,
\label{eq:AppGaussianDelay}
\end{equation}
i.e. about 20\% of the laser pulse duration due to the resonant excitation in the (1+2) REMPI scenario. 
\begin{table}[t]
	\begin{ruledtabular}
		\begin{tabular}{lcccc}
			& $\REMPIM=1$ & $\REMPIM=2$ & $\REMPIM=3$ & $\REMPIM=4$  \\
			\hline
			\hline
			$\REMPIN=1$ & 33.9 & 39.1 & 41.5 & 42.9\\
			$\REMPIN=2$ & {\bf 19.6} & 24.0 & 26.2 & 27.7 \\
			$\REMPIN=3$ & 13.8 & 17.5 & 19.6 & 20.9     \\
			$\REMPIN=4$ & 10.7 & 13.8 & 15.7 & 16.9
		\end{tabular}
	\end{ruledtabular}
	\caption{Resonant ionization delay in percent of the FWHM ($\Delta t$) for ($\REMPIM$+$\REMPIN$) REMPI with a Gaussian pulse.}
	\label{tab:AppDelayGaussian}
\end{table}

\subsection{Time-delays for ($\REMPIM$+$\REMPIN$) REMPI} \label{sec:App_AnaDelayGaussNM}
Generalizing the above procedure to ($\REMPIM$+$\REMPIN$) REMPI with a Gaussian pulse shape by replacing the resonant excitation field in Eq.~\eqref{eq:AppGaussian} by $\mathcal{E}^{\REMPIM}(t)$ and the ionizing field in Eq.~\eqref{eq:AppGaussianPhotoelectronAmplitude} by $\mathcal{E}^{\REMPIN}(t)$, we obtain a general expression for the resonant ionization delay
\begin{equation}
\tau^G_{(\REMPIM,\REMPIN)} = \frac{1}{\sqrt{\pi}} \, \sqrt{\frac{\REMPIM}{\REMPIN (\REMPIM+\REMPIN)}} \, \frac{\Delta t }{\beta_G}.
 \label{eq:AppGaussianNMdeleay}
\end{equation}
Specific values for resonant excitation with $\REMPIM=1,2,3,4$ photons and ionization with $\REMPIN=1,2,3,4$ photons are presented in Tab.~\ref{tab:AppDelayGaussian}. The value for the (1+2) REMPI scenario considered in the experiment is shown in bold. Eq.~\eqref{eq:AppGaussianNMdeleay} shows that the ionization delay increases with $\REMPIM$, i.e. the number of photons required for resonant excitation, and decreases with $\REMPIN$, i.e. the number of photons required for ionization. 

\subsection{Time-delays for REMPI with other pulse shapes}\label{sec:App_AnaDelayShape} 
\subsubsection{($\REMPIM$+$\REMPIN$)~REMPI with $\cos^2$ pulses}
Similar results, obtained for $\cos^2$ pulses defined as 
\begin{equation}
\mathcal{E}(t)= 
\begin{cases}
\mathcal{E}_0  \cos^2 \left( \beta_C \frac{t}{\Delta t} \right), & \text{if } -\frac{\pi \Delta t}{2 \beta_C} < t < \frac{\pi \Delta t}{2 \beta_C}	   \\
0,       & \text{otherwise}
\end{cases}
 \label{eq:AppCos2Field}
\end{equation}
with $\beta_C = 2 \arccos(2^{-\frac14})$ and an intensity FWHM of $\Delta t$, are shown in Tab.~\ref{tab:AppDelayCos2}.
\begin{table}[t]
	\begin{ruledtabular}
	\begin{tabular}{lcccc}
		& $\REMPIM=1$ & $\REMPIM=2$ & $\REMPIM=3$ & $\REMPIM=4$ \\
		\hline
		\hline
		$\REMPIN=1$ & 28.4 & 32.3 & 34.3 & 35.5\\
		$\REMPIN=2$ & {\bf 18.7} & 22.1 & 24.1 & 25.3\\
		$\REMPIN=3$ & 14.0 & 16.9 & 18.7 & 19.9   \\
		$\REMPIN=4$ & 11.2 & 13.7 & 15.3 & 16.5   
	\end{tabular}
	\end{ruledtabular}
	\caption{Resonant ionization delay in percent of the FWHM ($\Delta t$) for ($\REMPIM$+$\REMPIN$) REMPI with a $\cos^2$ pulse.}
	\label{tab:AppDelayCos2}
\end{table}  
The ionization time-delay for ($\REMPIM$+$\REMPIN$)~REMPI is described by
\begin{equation} \label{app:eqtau}
\tau^C_{(\REMPIM,\REMPIN)} 
= \frac{\Delta t}{\pi \beta_C} \left[ \zeta(\infty) - \zeta(\REMPIN) + \frac{1}{2} \zeta(\REMPIM) + \mathcal{S}_{\REMPIM,\REMPIN} \right],
\end{equation}
where the sum $\mathcal{S}_{\REMPIM,\REMPIN}$ is defined as 
\begin{equation}
\mathcal{S}_{\REMPIM,\REMPIN} =\frac{8 (\REMPIM!)^2 (\REMPIN!)^2 }{ (2\REMPIM)! (2\REMPIN)! } \sum_{p=1}^{\REMPIM} \sum_{q=1}^{\REMPIN} 
\binom{2 \REMPIM}{\REMPIM-p} \binom{2 \REMPIN}{\REMPIN-q} \, \sigma(p,q)
\label{eq:AppCos2DelayM+N2}
\end{equation}
with the terms $\sigma(p,q)$ 
\begin{equation}
\sigma(p,q)= 
\begin{cases}
\frac14 \, \frac{(-1)^{p+q}}{q^2 - p^2},		& \text{if} \quad p^2 \neq q^2 \\
-\frac{1}{16 p^2}     							& \text{if} \quad p^2 = q^2.
\end{cases}
\label{eq:s_pq}
\end{equation}
In Eq.~\eqref{app:eqtau},  $\zeta(n)$ is the $n$-th harmonic number of second order
\begin{equation}
\zeta(n) =\sum_{k=1}^{n} \frac{1}{k^2}, \quad \mathrm{with } \quad \zeta(\infty) = \frac{\pi^2}{6}.
\label{eq:Appzetadef}
\end{equation}
For single-photon excitation and $\REMPIN$-photon ionization, i.e. (1+$\REMPIN$)~REMPI, the above equation reduces to
\begin{equation}
\tau^C_{(1,\REMPIN)} = 
\left[\zeta(\infty) - \zeta(\REMPIN)
+ \frac{1+ 2\REMPIN}{2 (1+\REMPIN)^2}  \right] \frac{\Delta t}{\pi \beta_C}.
\label{eq:AppCos2Delay1+N}
\end{equation}
The ionization time-delay in (1+2)~REMPI considered in the experiment and highlighted in Tab.~\ref{tab:AppDelayCos2} results in
\begin{equation}
\tau^C_{(1,2)} = 
\left(\frac{\pi^2}{6} - \frac{35}{36} \right) \frac{\Delta t}{\pi \beta_C}
\approx 0.187\,\Delta t.
\label{eq:AppCos2Delay}
\end{equation}

\subsubsection{($\REMPIM$+$\REMPIN$)~REMPI with Lorentzian pulses}
The ionization time-delays in the ($\REMPIM$+$\REMPIN$) REMPI for pulses with a Lorentzian-shaped envelope described by
\begin{equation}
\mathcal{E}(t) = \frac{\mathcal{E}_0 }{1+ \left( \beta_L \frac{t}{\Delta t} \right)^2},
\label{eq:AppLorentzField}
\end{equation}
with $\beta_L = 2 \sqrt{ \sqrt{2} -1 }$ and an intensity FWHM of $\Delta t$, are described by 
\begin{equation} 
\tau^L_{(\REMPIM,\REMPIN)} =  
\frac{\Gamma(\REMPIM) }{\Gamma \left( \REMPIM -\frac12 \right) } 
\frac{\Gamma(\REMPIN-1) }{\Gamma \left( \REMPIN -\frac12 \right) } 
\frac{\Gamma \left(\REMPIM + \REMPIN - \frac32 \right) }{\Gamma \left( \REMPIM + \REMPIN -1 \right) } 
\frac{\Delta t}{\sqrt{\pi} \beta_L}
\label{eq:AppLorentzianNMdeleay}
\end{equation}
and exemplified in Tab.~\ref{tab:AppDelayLorentz}. In Eq.~\eqref{eq:AppLorentzianNMdeleay}, $\Gamma(\REMPIN)=(\REMPIN-1)!$ is the Gamma function. Due to the extended wings of the Lorentzian, the first moment diverges for $\REMPIN=1$. As a consequence, the spectral phase has a discontinuity at $\varepsilon=0$ so that the time-delay is not defined in accordance with the divergent term $\Gamma(\REMPIN-1)$. Physically, this is due to the slowly vanishing ionization rate at $t\gg\Delta t$, when the population of the $4p$-state $|c_{4p}(t)|^2$ is nearly constant but the ionization field $\mathcal{E}(t)$ is still non-vanishing. 
\begin{table}[t]
		\begin{ruledtabular}
			\begin{tabular}{lcccc}
				& $\REMPIM=1$ & $\REMPIM=2$ & $\REMPIM=3$ & $\REMPIM=4$ \\
				\hline
				\hline
				$\REMPIN=2$ & {\bf 24.7} & 37.1 & 41.2 & 43.3\\
				$\REMPIN=3$ & 12.4 & 20.6 & 24.0 & 26.0   \\
				$\REMPIN=4$ & 8.24 & 14.4 & 17.3 & 19.0   
			\end{tabular}
		\end{ruledtabular}
		\caption{Resonant ionization delay in percent of the FWHM ($\Delta t$) for ($\REMPIM$+$\REMPIN$) REMPI with a Lorentzian pulse.}
		\label{tab:AppDelayLorentz}
	\end{table}  
For single-photon excitation, i.e. (1+$\REMPIN$)~REMPI, the ionization time-delay reduces to 
\begin{equation} 
\tau^L_{(1,\REMPIN)} =  
\frac{1}{\REMPIN-1}
\frac{\Delta t}{\pi \beta_L}
\label{eq:AppLorentzianNM1deleay}.
\end{equation} 
resulting in 
\begin{equation}
\tau^L_{(1,2)} = 
\frac{\Delta t}{\pi \beta_L}
\approx 0.247\,\Delta t.
\label{eq:AppLorentzian12}
\end{equation}
for (1+2)~REMPI. Inspection of Tab.~\ref{tab:AppDelayLorentz} shows that the qualitative behavior of the resonant ionization delays for $\REMPIN>1$ is similar to the Gaussian and the $\cos^2$ cases.

\subsubsection{($\REMPIM$+$\REMPIN$)~REMPI with rectangular pulses}
In contrast, rectangular pulses with 
\begin{equation}
\mathcal{E}(t) = \mathcal{E}_0 \, \Pi(t),
 \label{eq:AppRectField}
\end{equation}
where the rectangular shape function $\Pi(t)$ is defined as
\begin{equation}
\Pi(t)= 
\begin{cases}
1,		& \text{if } -\frac{\Delta t}{2} < t < \frac{\Delta t}{2}\\
0,       & \text{otherwise}
\end{cases}
\label{eq:AppPiFunction}
\end{equation}
do not change their shape when the $n$-th power of the field is considered, since
\begin{equation}
\mathcal{E}^{n}(t) = \mathcal{E}_0^{n} \, \Pi^{n}(t) = \mathcal{E}_0^{n} \, \Pi(t).
\label{eq:AppRectFieldM}
\end{equation}
As a consequence, the resonant ionization delay for rectangular pulses
\begin{equation}
\tau^R_{(\REMPIM,\REMPIN)} = \frac{\Delta t}{6} \approx 0.167\, \Delta t
 \label{eq:AppRectDelay}
\end{equation}
of about 16.7\% is identical for all ($\REMPIM$+$\REMPIN$) REMPI processes. 

\bibliography{ultra_db.bib}

\end{document}